\documentstyle[12pt,fleqn]{article}
\renewcommand{\baselinestretch}{1.2}

\newcommand{\tr}{{\rm tr}\, }

\newcommand{\fg}[1]{\item {\label{#1}}}

\newcommand{\barr}{\begin{array}}
\newcommand{\bea}{\begin{eqnarray}}
\newcommand{\beq}{\begin{equation}}
\newcommand{\ear}{\end{array}}
\newcommand{\eea}{\end{eqnarray}}
\newcommand{\ceq}{\nonumber \\ & & }
\newcommand{\eeq}{\end{equation}}

\newcommand{\continue}{\nonumber \\ }

\newcommand{\spav}[1]{\parbox{1mm}{\vspace*{#1}}}

\newcommand{\ssty}{\scriptstyle}
\newcommand{\sssty}{\scriptscriptstyle}

\newsavebox{\ipiu}
\newsavebox{\imen}

\sbox{\ipiu}{$\ssty i \sssty +1$}
\sbox{\imen}{$\ssty i \sssty -1$}

\begin{document}

\begin{titlepage}
\hfill \today\\
\spav{2cm}\\
\centering \spav{1cm}\\
{\LARGE\bf
%%BeginAbstract
 Periodic Orbit Theory
%%EndAbstract
\\}
\spav{2cm}\\
{\large
%%BeginAbstract
Bruno Eckhardt
%%EndAbstract
 }\\
{\normalsize\em Fachbereich Physik der Philipps-Universit\"at,
\\Renthof 6, D-3550 Marburg   \\ }
\spav{2mm}\\
\vfill
Lecture notes for the\\
International School of Physics ``Enrico Fermi''\\
on Quantum Chaos\\
Varenna, Villa Monastero, 23 July -- 2 August 1991\\
\vfill
\spav{4mm}\\
\end{titlepage}
\setcounter{footnote}{0}

\tableofcontents
\newpage

\section{Introduction}
Semiclassical quantization can be understood as an interference phenomenon.
Wave fronts propagate along classical trajectories and build up eigenfunctions
if they interfere constructively. In integrable systems tori form a backbone
for all classically allowed motions and the conditions for
constructive interference yield the well known WKB quantization
rule\cite{Ber83,BT76,BT77}.
Furthermore, rather detailed information on wave functions, matrix elements
and selection rules can be derived\cite{Delos1,Delos2}.

It appears that for chaotic systems periodic orbits play a role similar
to that of tori in integrable systems\cite{Gutz90}. Gutzwillers
famous stationary phase analysis\cite{Gp} of the trace of Green's function
provides a link between the quantum spectrum and classical periodic orbits.
This link has been made explicit for hydrogen in a magnetic
field\cite{Wel,Wi87} and some molecules\cite{Piq90}:
a Fourier transform of the spectrum reveals
sharp peaks at the poriods of classical periodic orbits.
However, periodic orbits are much too numerous to provide a one-to-one
connection between individual paths and quantum eigenvalues\cite{Ber83}.
Mathematically, this is reflected in the inherent divergence of the formal
Gutzwiller trace formula\cite{EA89,SS89}.
Experiments on microwave resonances\cite{Stoe} show that such
relations hold for more general wave phenomena as well.

Methods to overcome such divergences have been developed in the
context of general dynamical systems\cite{PC89,AACI,AACII},
where invariant sets can be characterized by their periodic points.
At the heart of these developments has been the observation that classical
periodic orbits are strictly organized, both topologically and metrically,
and that this organization can be
exploited to rewrite ill behaved sums over periodic orbits
in a convergent form. The final result is simple and computationally efficient,
sometimes showing faster than exponential convergence\cite{CC90}.

At present, for a succesful application of the program, a symbolic
organization of the dynamics is necessary. Because of this requirement,
it has been carried out for a few systems only, most
notably 2-d billiards formed by three disks or four
hyperbolas\cite{CE89,GSprl,SS91}, the anisotropic Kepler
problem\cite{GSprl,TW91} and collinear Helium\cite{RTW91}.
These studies (many relevant contributions are collected
in a recent conference proceeding\cite{Cop})
have demonstrated that indeed the trace formula
does yield semiclassical approximations to the eigenvalues.
They have also shown that the convergence behaviour can be
improved dramatically by imposing a functional equation\cite{GSprl,SS91}.
The existence of this functional equation is suggested by properties
of an $S$-matrix approach to quantization\cite{Bogo91a,Bogo91b,DoSmi91}
and by the existence of a similar relation for
the Riemann zeta function\cite{BeKea90,Kea91,KeaVar}
and for Selberg zeta functions on surfaces of
constant negative curvature\cite{Hej,BV}, but it has not been possible to
derive it within the semiclassical approximation (compare the
discussion in Ref.~\cite{Mayer})

Another point discussed in these notes concerns the extension of
Gutzwiller's theory to include matrix elements as well\cite{Wil1,Wil2}.
In principle, this requires the full Green's function, which can be expressed
as a sum over recurrent orbits, i.e., orbits returning to their initial
positions in projections, not necessarily in full phase space\cite{Bogo89}.
It turns out\cite{EFMW91} that in the case of sufficiently
smooth operators, one can again arrive at expressions
involving periodic orbits only.  This then establishes a complete link
between experimentally accessible spectra and classical
periodic orbits.

The key technical step will be to express sums over periodic orbits
as products over periodic orbits. Such products are termed `dynamical
zeta functions'\cite{Rue}, in analogy to Riemann's zeta
function\cite{Titch,Edwards},
which can be written as an infinite product over prime numbers,
\beq
\zeta_R^{-1} (s) = {\prod_{primes} (1-p^{-s}) } \, .
\eeq
Dynamical zeta functions look similar, the product extending over
contributions from primitive periodic orbits (labelled $p$),
\beq
\zeta_{Dyn}^{-1} (s) = \prod_{p} (1-t_p(s)) \, .
\eeq
The fact is, though, that the objects one is interested in are the products
as given above, almost never the inverses thereof.
Thus there is an inverse relationship in the behaviour of the two
types of functions. For instance, in accord with one approach to the
Riemann hypothesis\cite{Edwards}, one would like to identify the zeros of
$\zeta_R(s)$ along the critical line $s=1/2 + i t$ with eigenvalues of a
quantum system. However, dynamical zeta functions $\zeta_D(s)$
have {\rm poles} rather than zeros at the positions of the eigenvalues.

The theory will be discussed for two degree of freedom systems only,
since they are the at present most interesting class.
In many cases, extensions to more degrees of freedom are possible.

The organization of the paper is as follows. In the next section, I will
summarize Gutzwiller's theory for the spectrum of eigenenergies
and extend it to diagonal
matrix elements as well. The derivation of the associated zeta function
is given (2.2) and the identification of suitable scaling
variables discussed (2.3).

In section 3 tools necessary for the organization of
chaos will be discussed:
symbolic dynamics (3.1), the connectivity matrix (3.3), the topological
zeta function (3.4) and general transfer matrices and zeta functions (3.5).
Although illustrated for the case of hard collisions in a billiard, the
symbolic dynamics can be extended to `smooth collisions' in smooth potentials
(3.2).

In systems with discrete symmetries, zeta functions factorize into
zeta functions on invariant subspaces. This symmetry factorization
and the associated reduction in symbolics is discussed in section 4.

The ideas developed here are illustrated for the example of a free
particle reflected elastically off three disks in section 5.
Methods to find periodic orbits (5.1), the convergence of the
trace formula (5.2), the semiclassical computation of
scattering resonances (5.3),
the convergence of the cycle expansion (5.4) and
methods to obtain eigenvalues of the bounded billiard (5.5) are
discussed.

The relevant parts of a classical periodic orbit theory are developed in
section 6.1, including a discussion of escape rates and the
Hannay-Ozorio de Almeida sum rule (6.2).

Finally, the issue of semiclassical matrix elements is taken up again
and applications to experiments are discussed.

\section{Semiclassical periodic orbit theory}
Following Gutzwiller\cite{Gutz90,Gp},
a connection between periodic orbits and quantum
properties can be derived from a stationary phase evaluation of
Feynman's path integral. Usually, only the trace of Green's function
is evaluated, but as will be shown below, a simple extension allows
for the computation of matrix elements as well.

\subsection{Expressions for $\tr GA$}

Starting point is the quantum mechanical expression for the trace of
Green's function times some observable, $g_A(E) = \tr GA$.
Expanded in the (complete) energy eigenbasis
(states $|n\rangle$ of energy $E_n$), this expression takes on the form
\beq
g_A(E) =  \lim_{\epsilon\rightarrow 0}
\sum_n {\langle n |A| n \rangle \over E - E_n+ i \epsilon}\, ,
\eeq
so that
\beq
\rho_A(E) = -{1 \over \pi} {\rm Im}\  g_A(E) =
\sum_n \langle n |A| n \rangle\, \delta( E - E_n)\, .
\label{q_tr}
\eeq
Thus $\rho_A$ has poles at the quantum eigenvalues, with residues given by
the matrix elements.

The way to obtain the semiclassical expression for (\ref{q_tr}) is to consider
a semiclassical approximation to the propagator, to Fourier transform to find
Green's function and then to take the trace.
Technically, since the observable can also depend on momenta, one
has to use a phase space representation such as
Wigner's function\cite{Wigner,EFMW91}.

Contributions to the trace come from two sources: from the very short paths,
where the propagator turns into a delta function\cite{BM72},
and from the periodic
paths. To evaluate the first part, one uses a Taylor series expansion of
the trajectory in powers of time $t$, and exploits the smallness of $t$ in
evaluating integrals\cite{BB72,GV}.
This then gives a smoothly varying contribution to $g_A$,
\beq
g_{A,0} = \int {d{\bf p} d{\bf q} \over h^N} \delta(E-
H({\bf p}, {\bf q})) A({\bf p}, {\bf q}) \, ,
\label{A_smooth}
\eeq
i.e.~the average of the observable over the energy shell.
$N$ is the number of degrees of freedom. Higher order corrections,
similar to the boundary and curvature corrections to Weyl's law for
billiards, can also be calculated\cite{BH,BB72,GV}.

The second part is obtained by approximating the propagator
as a superposition of contributions from all paths. After
a stationary phase evaluation of the Fourier transform one obtains
the oscillatory part of Green's function\cite{BM72},
\beq
G_{osc}({\bf q}', {\bf q}; E) =
        { 1 \over i \hbar (2 \pi i \hbar)^{(N-1)/2} }
\sum_{paths}  \left| D_S \right|^{1/2} e^{i S_p(E)/\hbar - i \pi \nu'_p /2}
\, ,
\eeq
where the sum extends over classical paths $p$ connecting ${\bf q}_1$ and
${\bf q}_2$ at a fixed energy $E$, irrespective of the time it takes;
$S_p$ is the classical action $\int {\bf p}\, d{\bf q}$,
\beq
D_S =
\det \left( \matrix{
{ \partial^2 S_p \over \partial {\bf q}' \partial {\bf q}} &
{ \partial^2 S_p \over \partial E \partial {\bf q}} \cr
{ \partial^2 S_p \over \partial {\bf q}' \partial E} &
{ \partial^2 S_p \over \partial E^2} \cr } \right)
\eeq
is the determinant of second derivatives and the index $\nu'_p$ counts
the number of caustics on the energy shell.

The next step is to take the trace of $GA$,
\beq
g_A = \int\, d^N {\bf q} \, G({\bf q}, {\bf q}; E) A({\bf q}) \, ,
\eeq
where for simplicity an observable depending on positions only has been
substituted.  In spirit with the
semiclassical nature of the entire calculation one would also like to
evaluate this integral in stationary phase. This is possible, if $A$
varies slowly on the scale of a wavenumber. The calculation then
continues very much as in Gutzwiller's case\cite{Gp,EFMW91}.

The phase is stationary if the
final and initial momenta coincide, which is the condition that the
trajectory be periodic. In the neighbourhood of every closed
path a coordinate system with $q_1$ along the path and $q_2,\dots, q_N$
perpendicular to it may be introduced.
Using the factorization of the determinant $D_S$ and the fact that
up to second order in the deviations from the trajectory the action only
depends on the stability matrix of the classical path,  one finds
\beq
{1 \over (2\pi i \hbar)^{(N-1)/2}} \int dq_2\cdots dq_N \, |D_{S_p}|^{1/2}
e^{i S_p({\bf q}) /\hbar - i \nu'_p \pi /2}
= {1 \over |\dot q_1|}
{ e^{i S_p /\hbar - i \mu_p \pi /2} \over | \det (M_p-1) |^{1/2} }.
\eeq
where $S_p$ is the action along the periodic orbit, $M_p$ is the stability
matrix around the orbit and the phase shift $\mu_p$ is the
Maslov index of the periodic orbit\cite{EW90}.

Since the stability matrix is independent of the position along the
path, there remains the integral
$\int dq_1 A({\bf q}) /\dot q_1$, which
by $dq/\dot q = dt$ may be written as a time integral over one period.
Allowing for multiple traversals of a periodic orbit, we finally
find for the contribution of one periodic orbit to $g_{A,osc}$,
\beq
g_{A,p} = {- i \over \hbar} A_p \sum_{r=1}^\infty
{ e^{(i S_p/\hbar - i \mu_p \pi/2)r} \over |\det(M_p^r-1)|^{1/2} }
\, ,
\eeq
with $A_p$ the integral of $A$ along the orbit.

Combining the smooth part (\ref{A_smooth})
with the contributions from all periodic paths, one finds
\bea
\rho_A(E) &=& - {1 \over \pi} {\rm Im\ } {\rm tr\,} g_A(E) =
         \rho_{A,0}(E) + \sum_p \rho_{A,p}(E) \,              \cr
        &=& \int { d{\bf q} d{\bf p} \over h^N} A({\bf q}, {\bf p})
           \delta(E-H({\bf q}, {\bf p})) \,                   \cr
 &\  & + {\rm Im\,} \> {i \over \pi \hbar}
\sum_{p} \sum_{r=1}^\infty { A_p  \over | \det (M_p^r-1) |^{1/2}  }
e^{(i S_p(E)/\hbar - i \pi \mu /2)r}
\label{dens}
\eea
where
\beq
A_p = \int_0^{T_p} dt\, A({\bf q}_p(t), {\bf p}_p(t)) \, .
\eeq
In the final formula general observables $A({\bf p}, {\bf q})$
have been admitted. The momentum ${\bf p}(t)$ is then fixed to be the
momentum along the path at ${\bf q}(t)$.
The average of the observable over one period of the classical
trajectory $({\bf p}(t), {\bf q}(t))$ is $A_p / T_p$.
The key requirement, beyond the
applicability of a semiclassical approximation, is that the
observable be sufficiently smooth. Otherwise a steepest descent
approximation to all integrals has to be used.
>From the above discussion (eq~(\ref{q_tr})) one expects
this expression to show poles at the (semiclassical)
eigenvalues, the residues being the matrix elements.

\subsection{Selberg's and other zeta functions}

For the density of states the operator $A=1$ and thus $A_p=T_p$.
Then the contributions from periodic orbits to (\ref{dens})
may be rewritten as the logarithmic derivative of
an infinite product over periodic orbits\cite{Vor},
similar to the Selberg zeta function\cite{Selberg}
in the theory of geodesic motion on surfaces of constant negative
curvature\cite{Hej}.

With $A=1$ and $A_p = T_p$ the period,
the contribution from periodic orbits to (\ref{dens}) can be written
\beq
R_{A,osc} = {i \over \pi \hbar}
\sum_p\sum_{r=1}^\infty { T_p \over |\det (M_p^r-1)|^{1/2}}
e^{(i S_p / \hbar - i \mu_p \pi / 2) r} \,.
\eeq
The degrees of freedom enter in the size of the linearization perpendicular to
the orbit. For a two degree of freedom system, $M_p$ is a $2\times2$
matrix of determinant one. If the orbit is unstable, the eigenvalues are
$\Lambda_p$ and $1/\Lambda_p$. The denominator can then be expanded in a
geometric series\cite{Mill},
\beq
|\det (M_p^r-1)|^{-1/2} = |\Lambda_p|^{-r/2} (1-1/\Lambda_p)^{-1} =
\sum_{j=0}^\infty |\Lambda_p|^{-r/2} \Lambda_p^{-jr} \, ,
\eeq
so that
\beq
R_{A,osc} = {1 \over \pi \hbar}
\sum_p  \sum_{j=0}^\infty \sum_{r=1}^\infty  T_p \left\lbrack
e^{i S_p /\hbar - i \mu_p \pi /2} |\Lambda_p|^{-1/2} \Lambda_p^{-j}
\right\rbrack^r \, .
\eeq
Upon summing on $r$ one finds
\beq
R_{A,osc} = {1 \over \pi \hbar}
\sum_p \sum_{j=0}^\infty { T_p t_p^{(j)} \over
1- t_p^{(j)} } \, ,
\label{log_der}
\eeq
where  $t_p^{(j)} =
e^{i S_p/\hbar - i\mu_p\pi/2} |\Lambda_p|^{-1/2} \Lambda_p^{-j}$.
Using the relation $T_p = \partial S_p / \partial E$, one can write the
quotient in~(\ref{log_der}) as a logarithmic derivative,
\beq
R_{A,osc} = - {1 \over \pi}
\sum_p \sum_{j=0}^\infty {\partial \over \partial E}
\log (1-t_p^{(j)}) \, ,
\eeq
so that finally
\beq
R_{A,osc} = - {1 \over \pi }
{\partial \over \partial E} {\rm log} Z(E)
\label{lnZ}
\eeq
with the Selberg zeta function\cite{Selberg,Vor}
\beq
Z(E) = \prod_{j=0}^\infty \prod_{p} (1-
 e^{i S_p /\hbar- i \mu_p \pi / 2} |\Lambda_p|^{-1/2} \Lambda_p^{-j} ).
\label{selberg_zeta}
\eeq
Deriving zeta functions for the other traces involving matrix elements
requires a little trick and will be postponed until later (section 7.1).

Depending on ones application, it sometimes is convenient to think of
(\ref{selberg_zeta}) as an infinite product of dynamical zeta functions,
$Z = \prod_j 1/\zeta_j$ with
\beq
1/\zeta_j = \prod_{p} (1-
 e^{i S_p /\hbar- i \mu_p \pi / 2} |\Lambda_p|^{-1/2} \Lambda_p^{-j})\,.
\label{zeta_j}
\eeq

The leading order term $1/\zeta_0$ is Gutzwillers original
approximation\cite{Gp,Mill}, obtained by
replacing the determinant in the denominator by the dominant eigenvalue.
The discussion of convergence (section 5.2)
will reveal that indeed this first term
is the dominant one. All other zeta functions converge absolutely
and therefore cannot give eigenvalues or resonances near the real
energy axis\cite{Ikawa}.

\subsection{Scaling variables}

In general, the actions of periodic orbits are complicated functions of
energy\cite{Bar1}.
If the Hamiltonian describing the system has scaling properties,
e.g. if it is a sum of squares of the momenta plus a homogeneous potential,
$V(\lambda {\bf x}) = \lambda^\kappa V({\bf x})$,
then by a virial theorem, the action scales with energy like
\beq
S(E) = {2 \kappa E_0 \over (2+\kappa)} T(E_0)
{\left( {E \over E_0} \right)}^{(2+\kappa)/2\kappa} \, ,
\eeq
where $E_0$ is some reference energy and $T$ the period.
It thus becomes linear in the variable
\beq
k = {\left( {E \over E_0}\right)}^{(2+\kappa)/2\kappa)} \, .
\eeq
In the limit of a billard, $\kappa\rightarrow \infty$, $k$ is
essentially the usual wavenumber. Because of the simple
linear scaling of actions with $k$, one can use a Fourier
transform in this variable to uncover the periodic orbit
structures\cite{BBft,Wi87}.

Such scaling Hamiltonians are exceptional. However, in the limit of small
$\hbar$ one can expand the action to first order in energy around a
reference energy $E_0$, viz. $S(E) = S(E_0) + T(E_0) (E-E_0)$.
In this limit it is possible to identify an energy
interval which is classically small (the properties of periodic orbits
change little) but semiclassically large (the interval contains many quantum
eigenvalues). Then approximately $E$ itself is a good scaling variable.
Formally, this is equivalent to consider the eigenvalues as a function
of $\hbar$ for fixed classical energy, as used e.g.~in the derivation of
the spectral statistics of integrable systems\cite{Pois}.

\section{Organizing chaos}
In this section I provide the necessary formal background on symbolic
dynamics, transfer matrices and cycle expansions.
These tools are important in developing the theory of zeta functions
and their cycle expansions. Eventually, one might hope to overcome these
limitations.

\subsection{Symbolic dynamics}
The paradigmatic example of randomness is a coin toss\cite{Ford}, which,
at least in principle, yields as its outcome a string of heads and
tails, with no correlations between consecutive events. Thus all
strings are possible and equally likely. In a chaotic dynamical system,
one can find similar behaviour\cite{Bowen,Sinai,Rue,Lanf}:
first dynamics is reduced to a discrete
map using a Poincar\'e surface of section.  Then certain regions in
this section are assigned `heads' and `tails'. Depending
on where a trajectory crosses the surface of section, it
will map out a string of heads and tails, and different
trajectories will map out different strings. As discussed in MacKay's
lecture, such a construction is generically possible in the vicinity of a
homoclinic crossing\cite{Smale,Moser}. Different from the ideal coin
tossing experiment, the dynamical coin is loaded: the probability of
occurence of a given symbol is determined by the dynamics and need not be
the same for all symbols.

For a certain class of systems it seems possible to extend this symbolic
dynamics to all relevant regions of phase space.
Specifically, for three or four disks
arranged in a plane so that all lines connecting any two disks
are possible (and not shaded by a third disk), one has a
unique labelling of {\em periodic} orbits by disk visitation
sequences\cite{Eck87,gaspard1} (see Fig.~\ref{traj}).
As the disks are moved closer together
to form a bounded system, orbits disappear because of shading by one of the
disks\cite{SchE}. Nevertheless, one still seems to be able to
label all trapped periodic orbits uniquely by a string of symbols.

Here we focus on the three disk system, which is somewhat simpler than
the four disk billiard relevant for hydrogen in a magnetic field\cite{EW90}.
In both systems, every trajectory can be labelled by the disk visitation
sequence. The set of labels assigned to the disks is called the {\em alphabet}
(here: $\{1,2,3\}$), any string formed from them a {\em word} (the trajectory
shown in Fig.~\ref{traj} could be labelled by the word $1231312$). Evidently,
a particle cannot bounce off the same disk twice, so that repetitions of the
same symbol are prohibited. This exclusion of $\cdots 11 \cdots$,
$\cdots 22 \cdots$ and $\cdots 33 \cdots$ is a typical example of a
{\em grammer rule}.

The way in which infinite sequences specify periodic orbits is reminiscent
of the same construction in the horseshoe map\cite{Smale,Moser}. In a typical
scattering experiment, the ingoing direction is fixed and the impact
parameter varied. Then there will be an entire interval of
impact parameters containing trajectories with the same collision future
for the next $n$ collisions. In Fig.~\ref{s_int} the interval in
impact parameter leading to collisions with disk $1$ is indicated.
If the next collision is specified as well, a subinterval is selected.
With increasing number of collisions, these intervals shrink
to a point: thus, there will be exactly one impact parameter with the
prescribed collision sequence. The past of the trajectory depends
on the ingoing direction. Repeating then the same procedure
for the angle rather than the impact parameter a unique value of
both impact parameter and angle will be identified.
This construction is very similar to the way in which strings
and orbits are associated in the Baker's map\cite{Bil}.

\subsection{Smooth collisions}
The previous discussion might seem confined to billiard models.
However, it should be clear that smoothing the discontinuity at the
boundaries of the disks a little bit
will not change the topology of short orbits. For instance, the potential
$V(x,y) = (x y)^{2/d}$ is equivalent to a billiard bounded by the
hyperbola $xy =1$ for $d=0$ and changes to the quartic oscillator
$x^2 y^2$ for $d=1$. Dahlqvist and Russberg\cite{DR1,DR2} have followed periodic
orbits from $d=0$ (where a code is known) to $d=1$ to establish a
symbolic coding for the above quartic oscillator.
Since the potential becomes more repulsive as $d$
approaches one, it is difficult to imagine that new orbits are born rather
than existing ones destroyed. However, this has to be
checked case by case and is not always obvious.

An alternative way has been proposed in Ref.~\cite{EW90,EW91}.
A characteristic feature of collisions is that a change
in orientation in a local coordinate system takes place.
One can think of defining the coordinate system using two
neighbouring trajectories with parallel velocities. During the collision,
they will cross in position space, causing a change in orientation of the
local coordinate system.

Related to this change in orientation are self conjugate points,
where neighbouring trajectories started with momentum slightly different from
the reference trajectory return to it.  Then an off diagonal matrix
element of the monodromy matrix after a full period $T$ vanishes,
\beq
\left( \matrix{ \delta x_\perp \cr \delta p_\perp } \right)
(T) = \left( \matrix{ m_{xx} & 0 \cr
                      m_{xp} & m_{pp} } \right)
\left( \matrix{ \delta x_\perp \cr \delta p_\perp } \right)(0) \, .
\eeq
Regions where one would identify a bounce are bounded by two such
conjugate points.
As demonstrated in Fig.~\ref{orbits} this method also works for orbits
which are very close in position space and where it is not immediately
obvious whether they undergo a collision when approaching the equienergy
contour or not.

For the computation of self conjugate points, one can use the
linearized equations of motion and the monodromy matrix ${\bf M}$.
For a two degree of freedom
system, this is a $4\times 4$ matrix, which can be obtained by integrating
$16$ first order differential equations. Two eigenvalues of ${\bf M}$ after
a full period are equal to one, due to the fact that both a shift
along the orbit and a shift out of the energy shell will be preserved.
The interesting part of ${\bf M}$ is the $2\times 2$ matrix ${\bf m}$
describing neighbouring trajectories in a plane perpendicular to the orbit
but on the energy shell. As worked out in Ref.~\cite{EW91}, it is possible
to introduce a coordinate system in which the trivial directions are
eliminated and in which closed equations for the $2\times 2$
submatrix ${\bf m}$ can be found.

An additional advantage of a definition of a symbolic code in terms
of self conjugate points is the close connection to semiclassics.
Since the propagator has an amplitude
proportional to $\sqrt{dq_\perp(T) / dp_\perp(0)}$, vanishing of the off
diagonal element also signals a break down of the semicassical
approximation, the presence of a caustic and a change of the Maslov phase.
This close connection between Maslov indices and the symbolic code
is important for the cancellation of terms in the cycle
expansion (section 5.4).

\subsection{Connectivity matrix}
Given the division of phase space into cells labelled by some alphabet
the dynamics enters in form of transitions between different cells.
The {\em connectivity} matrix encodes the information whether it is
possible to go from one cell to another or not.
In its simplest form, it is defined by
\beq
T_{i,j} = \left\{ {1 \atop 0}
\qquad { {\rm if\ transition\ from\ }j{\rm\ to\ }
i{\rm\ is\ possible} \atop
 {\rm if\ it\ is\ not\ possible \qquad\qquad\qquad} } \right. \, ,
\eeq
where the indices are letters from the alphabet.

For many applications, especially transfer matrices,
a generalization based on refinements of the cells is required. Cells are
subdivided and labelled according to the past of trajectories in the
subcells. All
trajectories which cross the surface of section in a region inside cell $i_1$,
sharing a common past of crossings at $i_2, i_3, \cdots, i_N$, define a
unique smaller cell which will be labelled $i_1i_2\cdots i_N$. When
iterated once, they will cross the surface in any one of the regions with the
label $i_0 i_1 i_2, \cdots i_{N-1}$, where $i_0$ can (in principle) be any
symbol from the alphabet: the very last symbol is dropped and a new
one added. The connectivity matrix generalizes to one indexed by words
$I=i_1 i_2 \cdots i_N$ and $J=j_1 j_2 \cdots j_N$.
Since in one iteration only one symbol is dropped and replaced by a new
one, the connectivity matrix can have entries at positions with
coinciding intermediates only, i.e.\ $T_{I,J} \ne 0$ for
$I=a i_1 \cdots i_{N-1}$ and $J=i_1 \cdots i_{N-1} b$ only.

The number $N_n$ of allowed strings of length $n$ is given by the trace of
the $n$-th power of the connectivity matrix,
\beq
N_n = \tr T^n\, .
\eeq
For example, for the case of a complete binary
code (symbols $0$ and $1$, no grammer rule), the transfer matrix is
\beq
\matrix{ & & \matrix{ 0 & 1  } \cr
T_{bin} = &
         \matrix{ 0 \cr 1} &
         \left( \matrix{1 & 1  \cr
                        1 & 1  \cr} \right)
} \, ,
\eeq
where the numbers outside the matrix indicate the symbols.
One can also define the larger connectivity matrix, acting on pairs of
symbols,
\beq
\matrix{ & & \matrix{ 00 & 01 & 10 & 11 } \cr
T_{bin} = &
\matrix{ 00 \cr 01 \cr 10 \cr 11 } &
         \left( \matrix{\ 1\  & \ 1\  & \ 0\  & \ 0\     \cr
                        0 & 0 & 1 & 1    \cr
                        1 & 1 & 0 & 0    \cr
                        0 & 0 & 1 & 1    \cr} \right)
} \, ,
\eeq
It is easily checked that both matrices yield the same strings.
The number of strings of length $n$ that can be formed is $N_n= 2^n$.
For example, at length $2$, the four strings $00$, $01$, $10$ and
$11$ are possible. Evidently, when periodically continued,
$00$ and $11$ correspond to the fixed points $0$ and $1$ of
length $1$ and $01$ and $10$ describe the same periodic string.
The {\em primitive period} $n_p$ of a periodic string is the length of
the shortest block from which it can be obtained. By cyclic permutation,
there are $n_p$ such blocks.

Thus the total number of strings of length $n$ can be decomposed
into the number $M_d$ of primitive strings of length $d$ dividing $n$,
\beq
N_n = \sum_{d|n} M_d \,.
\label{nppo}
\eeq
By M\"obius inversion\cite{HW}, one finds
\beq
M_d = {1 \over n} \sum_{d|n} \mu\left({n \over d}\right) N_d \, ,
\eeq
where the M\"obius function is defined by $\mu(1) = 1$, $\mu(n)=0$ if
$n$ contains the square of a prime and $\mu(n) = (-1)^k$
if $n$ contains $k$ prime factors.
Some examples are given in table~1.

In case of the three disks with their exclusion rule, the
associated connectivity matrix reads
\beq
\matrix{ & & \matrix{ 1 & 2 & 3 } \cr
T_{3-d} = &
          \matrix{ 1 \cr 2 \cr 3 } &
          \left( \matrix{0 & 1 & 1 \cr
                         1 & 0 & 1 \cr
                         1 & 1 & 0 \cr} \right)
} \, ,
\eeq
the number of strings is $N_n = 2^n + (-1)^n 2$ and the number
of primitive cycles agrees with the one for the binary code, except
for $n=1$ and $n=2$ (see table 1).

\subsection{Topological zeta function}

The number of strings of length $n$ is given by $\tr T^n$ and
thus dominated by the largest eigenvalue of $T$. The inverse of the largest
eigenvalue is a zero of $\det(1-zT)$. Using the identity
$\det A = \exp \tr \ln A$ and expanding the logarithm, one finds
\beq
\det(1-zT) =
\exp\left(-\sum_{n=1}^\infty {z^n \over n} \tr T^n \right) \, .
\label{d_et}
\eeq
If $T$ is the connectivity matrix, then $\tr T^n = N_n$. Using
the decomposition (\ref{nppo}) one can then replace the sum on $n$ by
one on all primitive periodic orbits $p$  of symbol length $n_p$
and their repetitions $r$,
\bea
-\sum_{n=1}^\infty {z^n \over n} \tr T^n
&=& - \sum_{p} \sum_{r=1}^\infty {z^{n_p r} \over n_p r} n_p  \cr
&=& + \sum_p \ln (1-z^{n_p}) \, ,
\eea
so that
\beq
\det (1-zT) = \prod_p (1-z^{n_p}) \, .
\label{d_pr}
\eeq
Such products over periodic orbits, formed in analogy to the Riemann zeta
function, are called dynamical zeta function\cite{Rue}, or,
if they are derived from
the connectivity matrix, topological zeta functions. As explained in the
introduction they are denoted by $1/\zeta$, although it is exactly the
product(\ref{d_pr}) and not its inverse which is studied.

For a complete code on $m$ symbols, the left hand side is easily evaluated
to be $1-mz$. Note the tremendous cancellations this must imply for the
infinite product on the right hand side when expanded as a power
series in $z$! For example, for a binary code, one has
\beq
1/\zeta_{bin} = 1-2z = (1-z)^2 (1-z^2) (1-z^3)^2 (1-z^4)^3 \cdots \,.
\eeq
This calculation can be used for the three disks as well. The number of
periodic orbits agrees with that for a complete binary coding except for
$n=1$, where there is no periodic orbit and $n=2$ where there are three
rather than just one. Therefore, the topological zeta function for three
disks can be obtained from that for the binary case,
\bea
1/\zeta_{3-d} &=& 1/\zeta_{bin} {(1-z^2)^2 \over (1-z)^2} \cr
              &=& (1-2z) (1+z)^2 = 1 - 3 z^2 - 2 z^3
\eea
which is still a finite polynominal.
If however just one orbit is missing, say one of the fixed points, then the
topological zeta function is no longer polynominal,
\beq
1/\zeta_{pr} = 1/\zeta_{bin} {1 \over 1-z} = 1 - z - z^2 - z^3 - \cdots
\eeq
Obviously, the leading zero and thus the topological entropy is still two.

\subsection{Transfer matrices and cycle expansion}

Transfer matrices have the same structure as connectivity matrices and the
same vanishing elements, but the $1$'s are replaced by quantities
multiplicative along trajectories. They provide the connection to
classical statistical mechanics\cite{Bowen,Sinai,Rue,Lanf} and have been
used in semiclassical mechanics first by Gutzwiller\cite{Gutz90,G_Tm1,G_Tm2}
in his analysis of the anisotropic Kepler problem. They also figure
prominently in Bogomolny's\cite{Bogo91a,Bogo91b}
theory of semiclassical quantization. Here, the precise form of
{\em off}-diagonal matrix elements is not so important, since all relevant
quantities will be expressed in terms of traces of powers of $T$,
which involve periodic orbits only. It should be noted that while there
is some ambiguity in assigning matrix elements of $T$, there is none
for periodic orbits and thus traces of $T^n$: actions, periods and
stability exponents are representation independent.

Entries of the transfer matrix or powers thereof
are labelled by the code of the initial and
final cells. Diagonal elements are thus associated with
contributions from trajectories that start in one cell and return.
An increased resolution with its longer code for the cells means that initial
conditions for trajectories returning to that cell have to be specified more
precisely, collapsing to a point in the limit of infinite resolution.
A primitive cycle of length $n$ will contribute $n$ times.
If $p\in(d)$ denotes the different primitive cycles of length $d$
and $t_p$ the contribution from cycle $p$, then
\beq
\tr T^n = \sum_{d|n} \sum_{p\in(d)} d\, t_p \, .
\eeq
Upon substitution into (\ref{d_et}) and manipulations similar to
the ones that lead to (\ref{d_pr}) one finds
\beq
\det(1-zT) = \prod_p (1-z^{n_p} t_p) \, .
\label{d_cyc}
\eeq
In general, the transfer matrix will depend on variables (such as the
wavenumber in case of the Gutzwiller trace formula) and one is interested
in the zeros of (\ref{d_cyc}) as a function of this variable.
Therefore, $z=1$. However, $z$ is extremely valuable as an auxiliary
variable when organizing the product. It is only for the final
calculation that one puts $z=1$.

The periodic orbits may conveniently be labelled by their symbolic
codes. For the case of a complete binary code one thus finds
\bea
\det(1-zT) & = &
   (1-zt_{0})(1-zt_{1})(1-z^2 t_{01})(1-z^3 t_{001})(1-z^3 t_{011}) \ceq
   (1-z^4t_{0001}) (1-z^4t_{0011})(1-z^4t_{0111})(1-z^5t_{00001})
            (1-z^5t_{00011}) \ceq
            (1-z^5t_{00101})(1-z^5t_{00111})(1-z^5t_{01011})(1-z^5t_{01111})
\dots \, .
\label{zetabin}
\eea
The cycle expansion is now obtained by factoring out the products and
arranging terms in a power series in $z$, just as in case of the
topological zeta function,
\begin{eqnarray}
1/\zeta &=&  1 - zt_0 - zt_1
 - z^2\lbrack (t_{01} - t_1 t_0) \rbrack    \nonumber\\
 & & - z^3\lbrack (t_{001}- t_{01} t_0)
     - (t_{011}- t_{01} t_1) \rbrack        \nonumber\\
 & & - z^4\lbrack (t_{0001}  - t_0 t_{001}) + (t_{0111} - t_{011} t_{1})
                                            \nonumber\\
& & + (t_{0011}  -  t_{001} t_{1} - t_{0} t_{011} + t_0 t_{01} t_1) \rbrack-
\dots \, . \\
\label{curvbin}
&=& \sum_n c_n z^n
\label{curv_n}
\end{eqnarray}
The important feature to note is that the contributions $t_0$ and $t_1$ from
the two fixed points stand isolated but that all others come in groups.
In the limit of $t_p \rightarrow 1$ the connectivity matrix is recovered and as
eq~(\ref{zetabin}) shows, the cancellations among all higher order terms are
complete. This is the main use of the topological zeta function in this
context: it provides a back bone of possible contributions to periodic
orbit expressions and shows how they are organized.
What remains to be checked is that the coefficients $c_n$
containing long periodic orbits can be grouped so that cancellations similar
to the ones for the topological polynominal actually take place.

Alternatively, one can start from (\ref{d_et}) and expand directly
in a power series in $z$, obtaining expressions reminiscent of a cumulant
expansion,
\beq
\det(1-zT) = 1 - z \tr T - {z^2 \over 2} (\tr T^2 - (\tr T)^2)
- {z^3 \over 3} ( \tr T^3 - \cdots) \cdots
\label{d_tr}
\eeq
For instance, the contributions to $\tr T^2$ are $(T^2)_{00,00} = t_{0}^2$,
$(T^2)_{11,11} = t_{1}^2$, $(T^2)_{10,10}=t_{10}$ and $(T^2)_{01,01}=t_{10}$.
Thus, the second term reduces to
\beq
{1 \over 2} (\tr T^2 - (\tr T)^2) = {1 \over 2} (t_0^2 + t_1^2 + 2 t_{01}^2
 -(t_0 + t_1)^2) = t_{01} - t_0 t_1
\eeq
in agreement with the cycle expansion (\ref{curvbin}).

For the case of three disks with the ternary alphabet with exclusion rules,
the zeta function is given by
\bea
1/\zeta & = & (1-z^2t_{12})(1-z^2t_{13})(1-z^2t_{23})
     (1-z^3t_{123})(1-z^3t_{132})
        \ceq
     (1-z^4t_{1213})(1-z^4t_{1232})(1-z^4t_{1323})(1-z^5t_{12123})\cdots
        \continue
     & = & 1 - z^2t_{12} - z^2t_{23} - z^2 t_{31}
     -  z^3 t_{123} -  z^3t_{132} \ceq
     -  z^4 \lbrack (t_{1213} - t_{12}t_{13})
     +  (t_{1232} - t_{12}t_{23})
     +  (t_{1323} - t_{13}t_{23}) \rbrack
         \ceq
     -  z^5 \lbrack ( t_{12123} - t_{12} t_{123}) + \cdots\rbrack - \cdots
\, .
\label{3d-zet}
\eea
Again the terms that stay isolated are exactly the ones indicated by the
topological zeta function.
If the disks are place in a symmetric arrangement then there are relations
between the orbits and the zeta functions factorize and simplify in a
beautiful manner to be explained in the next section.

\section{Symmetries}

Many dynamical systems of interest come equipped with symmetries.
Continuous symmetries usually give rise to conserved quantities by Noether's
theorem. Discrete symmetries provide relations between trajectories and can
be used to decompose phase space and dynamics into irreducible subspaces,
just as in the familiar case of a quantum system with symmetry where
eigenvalues and eigenvectors can be determined for the invariant subspaces
separately\cite{Gutz90,Rob,Laur,CE89,CESym}.

Basic to this is the observation that a discrete symmetry can act on an
orbit in two ways: it can map the set of points making up the orbit into
itself or it can map it into a different set which then again is an orbit.
In the latter case, the properties (actions, periods, stabilities) of the
orbit are unchanged, so that some factors in (\ref{3d-zet}) coincide.

The symmetry group of three circular disks arranged on the vertices of an
equilateral triangle is $C_{3v}$, consisting
of the identity $e$, two rotations $C_3$ and $C_3^2$ by $2\pi/3$ and $4\pi/3$
around the center and three reflections $c_{12}$, $c_{13}$ and $c_{23}$ on
symmetry lines (see Fig.~\ref{3d_symm}).
For instance, the rotations map the orbit $\overline{12}$
into $\overline{13}$ and $\overline{23}$. Any one reflection maps
$\overline{123}$ into $\overline{132}$ and so forth.
Taking just one representative of every degenerate class of orbits,
the zeta function (\ref{3d-zet}) becomes
\bea
1/\zeta & = & (1-z^2t_{12})^3 (1-z^3t_{123})^2 (1-z^4t_{1213})^3 \ceq
    (1-z^5t_{12123})^6 (1-z^6t_{121213})^{6}(1-z^6t_{121323})^3  \dots
        \continue
     & = & 1 - 3z^2\, t_{12} - 2z^3\, t_{123} - 3z^4\, (t_{1213} - t_{12}^2)
     - 6z^5\, ( t_{12123} - t_{12} t_{123}) \ceq
     - z^6\,( 6\, t_{121213} + 3\, t_{121323} +  t_{12}^3
                - 9\, t_{12} t_{1213} - t_{123}^2 )  \ceq
     - 6z^7\,( t_{1212123} + t_{1212313} + t_{1213123}
               + t_{12}^2 t_{123}  - 3\, t_{12} t_{12123}
               - t_{123} t_{1213}  ) \ceq
     -  3z^8\,( 2\, t_{12121213} + t_{12121313}
                  + 2\, t_{12121323} + 2\, t_{12123123} \ceq
                 ~~~~~ + 2\, t_{12123213} + t_{12132123}
        + 3\, t_{12}^2 t_{1213} + t_{12} t_{123}^2 \ceq
       ~~~~~- 6\, t_{12} t_{121213} - 3\, t_{12} t_{121323} -
4\, t_{123} t_{12123} - t_{1213}^2) - \cdots
\label{3dzeta}
\eea

A further reduction may be achieved by considering orbits whose
trajectories are mapped into themselves under a symmetry operation.
Then the orbit can actually be subdivided into irreducible
segments, the full orbit being a combination of several segments.
Similarly, the plane may be divided into a fundamental
domain(Fig.~\ref{3d_symm}) and its images under the symmetry operations.
One can then define a new code based on the group elements needed to map a
trajectory back onto the fundamental domain.
The fundamental operation is a reflection every time the particle hits
the boundary of the fundamental domain. If this is the only reflection
between two collisions with the disk, a symbol $0$ is assigned,
but if two are needed (corresponding to a rotation), then the symbol
is $1$. This new code turns out to be binary without any restrictions.
In Table~2 some orbits and their binary and ternary codes are listed.

For instance, the orbit $\overline{123}$ is invariant under the
rotations $C_3$ and
$C_3^2$. It can be pieced together from three identical segments $12$,
$23$ and $31$, mapped into each other by a rotation.
Under a reflection, this orbit goes over into $\overline{321}$, which
is just the time reversed orbit and therefore has the same symmetry.
Its contribution to the zeta function can thus be written
\beq
(1-t_{123})^2 = (1-t_{1}^3)^2 \, ,
\eeq
where the new label $1$ comes from the fact that the orbit can be mapped
back into the fundamental domain by a rotation.
Similarly, the orbit $\overline{12}$ is invariant under the reflection
$\sigma_{12}$, so it can be pieced together from two segments $12$ and
$21$. Application of the rotation
produces two more orbits $\overline{23}$ and $\overline{31}$.
Its contribution to the zeta function can thus be written
\beq
(1-t_{12})^2 = (1-t_{0}^2)^3 \, .
\eeq
Then there are orbits without any symmetry relations, which have
multiplicity $6$. Finally, there are orbits related by time reversal symmetry
but no other geometrical symmetry--only one member needs to be computed,
which then enters with multiplicity $12$.

The transfer operator is a linear operator and can therefore be decomposed
into a direct sum of its irreducible representations, implying a
factorization of zeta functions into products of zeta functions for
the irreducible subspaces. An explicit construction of the transfer
matrix based on the irreducible segments of an orbit is possible,
but not necessary. Of interest are determinants, for which there is
an expression involving traces only,
\bea
\det (1+M) &=& 1 + \tr M + {1 \over 2}\left( (\tr M)^2 - \tr M^2 \right)
\ceq
 + {1 \over 6} \left( (\tr M)^3 - 3 \, (\tr M) (\tr M^2) + 2 \,\tr M^3\right)
\ceq
+ \cdots + {1 \over d!} \left( (\tr M)^d - \cdots \right)
\,\, .
\label{factzet}
\eea
$d$ is the dimension of the representation. Since $M$ is essentially
a matrix representation of the group element under which the orbit
is invariant, its traces are given by the
characters $\chi_{\alpha}({\bf g})= \tr D_{\alpha}({\bf g})$,
listed  in standard tables\cite{hamer}. In terms of characters, we then have
for the 1-dimensional representations
\beq
 \det(1-D_{\alpha}({\bf g}) t) = 1 - \chi_{\alpha}({\bf g}) t
\,\, ,
\eeq
and for the 2-dimensional representations
\beq
\det(1-D_{\alpha}({\bf g}) t) = 1 - \chi_{\alpha}({\bf g}) t +
{1 \over 2} \left(
\chi_{\alpha}({\bf g})^2 - \chi_{\alpha}({\bf g}^2)\right) t^2 .
\eeq

Specifically, for the case of three symmetrically arranged disks and
$C_{3v}$ symmetry, one has two one-dimensional irreducible representations,
symmetric and antisymmetric under reflections, denoted $A_1$ and $A_2$, and two
degenerate two-dimensional representations of mixed symmetry, denoted $E$.
The contribution of an orbit with symmetry $g$ to the $1/\zeta$ Euler product
(\ref{factzet}) factorizes according to
\bea
\det(1-D({\bf g})t) &=& \left(1 - \chi_{A_1}({\bf g}) t \right)
\left(1 - \chi_{A_2}({\bf g}) t \right)
\left(1 - \chi_{E}({\bf g}) t + \chi_{A_2}({\bf g}) t^2 \right)^2
\,\, .
\label{fact3d}
\eea

Using the character table for the $C_{3v}$ group,
\vskip .3cm
\begin{center}
\begin{tabular}{|c|crr|}
\hline
$C_{3v}$     & $A_1$& $A_2$& $E$ \\
\hline
$ e   $      &   1  &   1  &  2  \\
$ C_3,C_3^2$ &   1  &   1  & $-1$  \\
$ \sigma_v$  &   1  &  $-1$  &  0  \\
\hline
\end{tabular}
\end{center}
\vskip .3cm
\noindent
one finds the following contributions from cycles:
\bea
  {\bf g}_{\tilde{p}}
&\ & \quad A_1 \quad \quad A_2 \quad \quad E \continue
e: \quad
( 1 - t_{\tilde p} )^6 &  = & (1 - t_{\tilde p})(1 - t_{\tilde p})
  (1 - 2 t_{\tilde p} + t_{\tilde p}^{2})^2 \continue
%             {\bf g}_{\tilde{p}} =
C_3,C_3^2: \quad
( 1 - t^3_{\tilde p} )^2 & = & (1 - t_{\tilde p}) (1 - t_{\tilde p})
                (1 + ~t_{\tilde p} + t_{\tilde p}^{2})^2 \continue
%             {\bf g}_{\tilde{p}} =
\sigma_i: \quad
( 1 - t^2_{\tilde p} )^3 & = & (1 - t_{\tilde p}) (1 + t_{\tilde p})
                (1 + 0 t_{\tilde p} - t_{\tilde p}^{2})^2 \, ,
\label{symm}
\eea
where $\tilde{p}$ denotes the symmetry reduced binary code for the
segments of the orbit.

The outcome of this exercise is that the factorization within the $A_1$
subspace is given by that of the binary zeta function (\ref{curvbin}),
and that the one for the
$A_2$ case is similar, except that the contributions from orbits with an
odd number of $0$'s change sign.
More interesting is the zeta function for the $E$ subspace, which contains a
different pattern of terms,
\bea
1/\zeta_E & = & (1+zt_{1}+z^2t_{1}^2)(1-z^2t_{0}^2)
                (1+zt_{100}+z^2t_{100}^2)  (1-z^2t_{10}^2) \ceq
                (1+zt_{1001}+z^2t_{1001}^2)
                (1+zt_{10000}+z^2t_{10000}^2) \ceq
                (1+zt_{10101}+z^2t_{10101}^2) (1-z^2t_{10011})^2
                 \dots  \continue
        & = & 1+z t_{1}+z^2(t_{1}^2 -t_{0}^2)
               +z^3 (t_{001}-t_{1} t_{0}^2) \ceq
        + z^4 \left[ t_{0011}+(t_{001} -t_{1} t_{0}^2  ) t_{1}
          -t_{01}^2 \right]    \ceq
         + z^5 \left[t_{00001}+t_{01011}-2 t_{00111}
         +(t_{0011} - t_{01}^2) t_{1}
  +( t_{1}^2-t_{0}^2) t_{100} \right]
  + \cdots
\label{3dzetam}
\eea
 Similar decompositions hold for other symmetry groups\cite{Laur,CESym}.

\section{The three disk system}
The formal developments of the previous sections will now be applied to the
three disk billiard\cite{Eck87,CE89,SchE,gaspard1,gaspard23}. This system
is ideally suited to test the methods since a good symbolic dynamics
is known and tuning of a parameter allows one to study the transition from
an open strongly chaotic system to a bounded one.

\subsection{Periodic orbits}
The example we will consider is motion of a point particle in the plane
with three circles removed. The system is characterized by the ratio
$d/R$ of the distance $d$ between the centers of the disks and their
radius $R$. If the distance between the circles is larger than the radius,
$d/R>2$,  then all of the plane is classically accessible and we have
a scattering geometry. If on the other hand the three disks touch then they
enclose a tipped region which we will refer to as the bounded billiard.

The classical dynamics of this system reduces to an exercise in geometry.
Several coordinate systems are possible: either position along the
circumference of the disks and parallel momentum or position and length of
the segment between any two collisions\cite{gaspard1,SSBil}
or scattering coordinates, i.e.\ impact parameter and ingoing angle.
More important is the choice of a numerically stable routine to find
the orbit\cite{SSBil,Bar2}. Drawing on general experience in numerical
mathematics a multipoint shooting method suggests itself. All intermediate point
trajectory are allowed to vary, so that for an orbit of symbol length $n$
one has $2n$ variables. A Newton-Raphson iteration will typically converge
very rapidly.

\subsection{Convergence of the trace formula}

The exponent in the semiclassical expression for the density of
states (\ref{dens}) becomes for billiards $S(E)/\hbar = L k$ with
$k = \sqrt{2 m E}/\hbar$ the wavenumber and $L$ the geometrical length
of the paths. All lengths can be taken relative to the radius of the disks.
The semiclassical limit $\hbar\rightarrow 0$ corresponds to $k\rightarrow
\infty$.

In many systems, the number of periodic orbits in (\ref{dens}) increases and
their weight decreases exponentially with period. The balance between these
effects determines con\-vergence\cite{EA89,SS89}.
For billiards, the period of an orbit is proportional to its length,
so that one has equivalent statements for the proliferation of orbits
with increasing geometrical length. The absolute convergence
of the series (\ref{dens}) is determined by the sum over absolute values.
Allowing for complex wavenumbers $k=k_r +i s$, this becomes
\beq
\sum_{p} \sum_{r=1}^\infty
{L_p \over |\det(1-M_p^r)|^{1/2}} e^{- r L_p s }\,.
\eeq
This expression will converge for sufficiently large $s$, but will diverge
for small $s$. As we are interested in the asymptotic behaviour for
long orbits, we can replace $\det(1-M_p^r) \approx |\Lambda_p|^r$,
where $|\Lambda_p|$ is the expanding eigenvalue of $M_p$.
Then
\bea
\sum_{p} \sum_{r=1}^\infty
 L_p  |\Lambda_p|^{-r/2} e^{- r s L_p }
 &=& \sum_{p} L_p {|\Lambda_p|^{-1/2} e^{- sL_p} \over
       1 - |\Lambda_p|^{-1/2} e^{- s L_p  } }                   \cr
 &=& \sum_{p} {\partial \over \partial s} \log
      \left( { 1 - |\Lambda_p|^{-1/2} e^{-s L_p } } \right)   \cr
 &=& {\partial \over \partial s} \log \zeta^{-1}(s)
\label{z_deriv}
\eea
with the zeta function
\beq
1/\zeta =  \prod_{p}
 \left( { 1 - |\Lambda_p|^{-1/2} e^{-s L_p } } \right) \, .
\label{zet_conv_qm}
\eeq
This is yet another example of a dynamical zeta function, where the weights
assigned to an orbit are $t_p = |\Lambda_p|^{-1/2} e^{-s L_p}$.
The abscissa of convergence then emerges as a zero of this zeta function.

Fig.~\ref{3-d_conv} shows $-s_0$, the negative of the abscissa of absolute
convergence. This representation was chosen because of the similarity between
$-s_0$ and the classical escape rate $\gamma$, to be discussed in
section 6.2. One curve was calculated using only the contributions from the
fixed points in a cycle expanded form of (\ref{zet_conv_qm}),
i.e.\ it shows the zero of
\beq
1 - |\Lambda_0|^{-1/2} e^{- s L_0} - |\Lambda_1|^{-1/2} e^{-s L_1} \, .
\eeq
For the second curve the contributions
\beq
- |\Lambda_{01}|^{-1/2} e^{-s L_{01}} + |\Lambda_0 \Lambda_1|^{-1/2}
e^{-s (L_0 + L_1)}
\eeq
from the period two orbit have been included. Both curves agree well for
$d/R>5$. For sufficiently separated disks the abscissa of absolute convergence
is negative, indicating convergence of the trace formula along the real axis.
Near $d/R \approx 2.8$ the curve crosses the real axis and approaches
$\approx 1.5$ for the closed billiard $d/R\rightarrow 2$.

The higher order zeta functions $1/\zeta_j$ in the Selberg Zeta function
contain more powers of $\Lambda^j$, so that their abscissa of convergence
satisfy $s_0^{(j)} < s_0^{(j-1)} < \cdots < s_0$. Since none if these functions
can have zeros in the half plane ${\rm Im}\, k > s_0^{(j)}$, the
resonances closest to the real axis in the interval
$s_0^{(1)} < {\rm Im}\, k < s_0^{(0)}$ come from $1/\zeta_0$ alone.
If $s_0^{(0)}<0$, that is for $d/R>2.8$ then
all resonances are semiclassically bounded away from the real axis.
This gap has previously been identified by Gaspard\cite{gaspard23}.

\subsection{Scattering resonances}

Scattering processes can be described using the $S$-matrix.
For the three disk system, the explicit expressions of the outgoing
waves in terms of the ingoing waves using the scattering matrix
have been given by Gaspard\cite{gaspard23}.
Resonances are related to complex zeros of
$\tr S^\dagger {dS \over dE}$, which is the extension of the density of
states to scattering systems\cite{BB74}.
It therefore is given by Gutzwiller's sum over classical periodic orbits,
eq~(\ref{dens}) with $A=1$.

As discussed in section 4 on symmetries, the Selberg zeta function factorizes
into three infinite products for the three subspaces of the symmetry group
$C_{3v}$ of the three disks. In the symmetry reduced symbolic code,
the cycle expansion for the $A_1$ subspace is given by the cycle
expansion for a complete binary code, eq~(\ref{curvbin}).
Using all periodic orbits up to symbol length 5, in total 14 orbits, one finds
the resonance spectrum (complex zeros) shown in Fig.~\ref{3-d-res}.
Evidently, the semiclassical calculations
agree well with the quantum results. Also, allmost all resonances are below the
limit\cite{gaspard23} predicted from the semiclassical calculations.
That the two resonances with smallest real part are above the semiclassical
limit is one of the deviations to be expected in the deep quantum regime.

\subsection{Convergence of the cycle expansion}

In recent years beautiful arguments for the convergence of the cycle
expanded zeta function in the realm of semiclassics have been developed.
Most important are attempts to obtain quantization conditions from a
scattering matrix\cite{Bogo91a,Bogo91b,DoSmi91}. The basic observation
is that for finite Planck's constant only a few scattering channels
are open so that upon neglect of evanescent waves one deals with
a finite scattering matrix. Therefore, the expansion of $\det(1-S)$
in a series in $\tr S^n$ as in eq~(\ref{factzet})
terminates. A different argument\cite{BeKea90,Kea91}
speculates about the existence of a Riemann Siegel relation and a
bootstrapping of the contributions from longer orbits. A Riemann
Siegel relation also holds within the scattering matrix formulation.

However, from a general point of view, the cycle expanded product should
converge because of the compensations induced by approximations of
long orbits from short ones\cite{PC89,AACI,AACII}.
In its most primitive form, the compensation argument applies to
terms in eq~(\ref{curvbin}) of the form
$t_{a^kb} - t_a t_{a^{k-1}b}$, involving a long orbit $a^kb$ and two
approximands $a$ and $a^{k-1}b$. This orbit is a periodic approximation
to an orbit homoclinic to $a$. Substituting the form of $t_{a^kb}$
from eq~(\ref{zeta_j}) for $j=0$, one finds
\beq
t_{a^kb} - t_a t_{a^{k-1}b} = t_{a^kb} \left( { 1-
e^{i (S_a + S_{a^{k-1}b} - S_{a^kb})/\hbar}
e^{- i \pi (\mu_a + \mu_{a^{k-1}b} - \mu_{a^kb})/2}
\left|{ \Lambda_a \Lambda_{a^{k-1}b} \over \Lambda_{a^kb} }\right|^{-1/2}
} \right) \, .
\eeq
Because of our definition of the code in terms of selfconjugate points
(section 3.2) the Maslov indices cancel, $\mu_a+\mu_{a^{k-1}b}-\mu_{a^kb}=0$.
Furthermore, since with increasing $k$, segments of $a^kb$ become closer
to $a$, the differences in action and the ratio of the
eigenvalues converge exponentially with the eigenvalue of the orbit $a$,
\bea
S_a + S_{a^{k-1}b} - S_{a^kb} &\approx& const\, \Lambda_a^{-k} \\
\left| \Lambda_a \Lambda_{a^{k-1}b} / \Lambda_{a^k b} \right|
 &\approx& \exp(-const\, \Lambda_a^{-k} )
\eea
Expanding the exponentials one thus finds that this term in the
cycle expansion is of the order of
\beq
t_{a^kb} - t_a t_{a^{k-1}b} \approx const\, \Lambda_a^{-k} \, .
\eeq
The number of terms in every order of the cycle expansion is even
larger than the number of periodic orbits\cite{CESym}. However, the
compensations reduce the size of the contributions from the periodic orbits,
inducing convergence.
In the case of the three disks, compensations should be best for $k=0$
since one then does not have to worry about phases. As Fig.~\ref{cycle_conv}
shows, compensation is very good for $d/R=6$ and $d/R=3$. What changes
as one approaches the closed billiard $d/R=2$ is that the convergence
is no longer as rapid and that pruning and missing orbits cause
non monotonic variations (see below).

The above discussion also shows that with increasing
energy more orbits are needed to obtain convergence: since the
actions increase with energy, so do their differences. But if the differences
become larger than $O(\hbar)$, it is no longer permitted to expand the
exponentials and these terms have to be kept. However, for sufficiently long
orbits, the compensations will again take place.

\subsection{Eigenvalues for the bounded three disk billiard}

Turning to the bounded billiard, one has to worry about additional problems
associated with pruning of orbits. As the disks are moved closer, some
orbits annihilate and dissapear\cite{SchE}. For instance the orbits
$\overline{000001}$ and $\overline{0000011}$ exist for
$d/R > 2.016...$ only. Other orbits in the family $\overline{0^n1}$
and $\overline{0^n11}$ with $n>5$ vanish earlier. In addition to these,
the orbit $\overline{0} = \overline{12}$ vanishes when the disks touch.
These are the only pruning rules for orbits of period $<10$.

As the disks move closer together, one would expect the
resonances to move up towards the real axis, lining up on it for the closed
system. The numerical calculations show this behaviour only approximately.
Including all orbits up to length 10 one finds that the resonances lie near
the real axis, but not exactly on it\cite{SchE}.
To remedy this, one can speculate about the existence of a functional
relation for the Selberg zeta function which would put it on the real
axis\cite{Bogo91a,Kea91}.

The existence of a functional equation is suggested by
the behaviour of the quantum mechanical integrated density of states
$N(E)$: it is a piecewise constant function that jumps by 1 at
every eigenvalue. To derive a semiclassical expression for this
note that the smooth density of states in eq (\ref{dens})
can be written as the logarithmic derivative,
\beq
\rho_0(E) = - {1 \over \pi}  {\partial \over \partial E} \log
\left( e^{- i \pi N_0} \right) \, ,
\eeq
with
\beq
N_0(E) = \int dE \, \rho_0(E)
\eeq
the integrated smooth density of states. Together with the
Selberg Zeta function (\ref{lnZ}), (\ref{selberg_zeta}) one finds
\beq
\rho(E) = - {1 \over \pi}  {\partial \over \partial E}
\log \left({  e^{- i \pi N_0} Z(E) } \right)
\eeq
or for the integrated density of states
\beq
N(E) = - {1 \over \pi}  \log \left( { e^{- i \pi N_0} Z(E) }\right)
\eeq
A piecewise constant function can be obtained if
the argument of the logarithm is real. Then $N(E)$ is constant inbetween
zeros and jumps by one, if the argument has a simple zero,
for then the phase jumps by $\pi$. Therefore, semiclassical
approximations to eigenvalues can be obtained as the zeros of
\beq
D(E) = { e^{- i \pi N_0} Z(E) } \, .
\label{f_det}
\eeq
This expression is very close to the functional determinant $\det(E-E_i)$,
(see, e.g., Ref.~\cite{VorC}).

For the billiard, the mean integrated density of states is given by the Weyl
expansion\cite{BH}, containing the
area, circumference and curvature of the billiard plus a term from the
singular tip at the touching point of the disks.
The contributions from the periodic orbits to the density of states
in the $A_1$ symmetry can be expanded as in eq~(\ref{curvbin}).
Fig.~\ref{DE} shows two approximations to $D(k)$ vs. $k$ for the three disk
billiard, one involving all orbits up to length 2 and length 3, respectively.
The overall agreement is good and most eigenvalues are resolved.
Again there are (quantum) deviations for small $k$.

\section{Classical periodic orbit theory}

When transferred to classical mechanics, the manipulations that lead from the
quantum propagator to the response function yield an expression for the
classical spectrum in terms of periodic orbits\cite{CEcl}. This is useful in
interpreting resonances in classical correlation functions. In addition, these
classical expressions contain information on classical escape rates in open
systems\cite{TK} and the Ozorio de Almeida-Hannay sum rule in bounded
systems\cite{OdAH}.

\subsection{The classical trace formula and associated zeta function}
Without much difficulty one can carry over Gutzwiller's
arguments\cite{Gutz90,Gp} from quantum mechanics to classical mechanics and
derive a periodic orbit representation for the classical propagator.
Starting point is Liouville's equation for phase space densities $f$,
\beq
\dot f = - i {\cal L} f \, ,
\eeq
with the formal solution
\beq
f_t = e^{- i {\cal L} t } f_0 \, .
\eeq
The equivalent of a Green's function may be defined by Fourier transform in
time,
\beq
G_{cl} = \int dt\, e^{i \omega t} e^{- i {\cal L} t } \, ,
\eeq
and the trace of Green's function then gives the density of states,
\beq
\rho_{cl} = \tr G_{cl} = \sum_i {1 \over \omega - \omega_i} \, .
\eeq
The spectrum of an ergodic system\cite{AA,Korn} has one eigenvalue at
$\omega=0$, connected with the invariant measure, superimposed on a continuum.
This continuum will show resonances which can be thought of as
complex poles $\omega_i$.

The calculation proceeds from an expression of the classical propagator as
an integral operator,
\beq
f_t({\bf x}) = \int d{\bf q} \, \delta({\bf x} - {\bf q}_t) f_0({\bf q})
\, ,
\eeq
where ${\bf q}_t$ denotes the point ${\bf q}$ propagated forward in time.
Upon taking the trace, the delta function singles out orbits that return to
their starting point: periodic orbits. Thus the delta function only gives a
contribution whenever $t$ equals a period $T_p$. Changing to a coordinate
system along the orbit and perpendicular to it one can do the perpendicular
integration, resulting in a factor $1/\det(1-M_p)$, and the integration along
the orbit, resulting in the period. In view of the delta function the
Fourier transform in time is easily done, leading to an expression
\beq
\rho_{cl} = \sum_{p} \sum_{r=1}^\infty
{ T_p \over \det(1-M_p^r) } e^{i r\omega T_p} \, ,
\eeq
which by manipulations similar to the ones in section 2.2
can be expressed as the logarithmic derivative of the zeta function
\beq
Z_{cl}(\omega) = \prod_{k=0}^\infty \prod_{p}
(1- |\Lambda_p|^{-1} \Lambda_p^{-k} e^{i \omega T_p})^{k+1} \, ,
\eeq
where the origin of the power $k+1$ may be traced to the fact that in
contrast to the quantum case one has a full determinant
rather than a square root in the denominator.

\subsection{Escape rates and sum rules}

I shall not give a complete discussion of the classical periodic orbit
expression and its use in determining resonances\cite{Rueres,CPR},
except for a discussion of the significance of a zero at real $s$.

Consider first a scattering system such as the arrangement of three
non-touching disks. One can then ask for the time a particle
will spend in the central region\cite{TK}. For this hyperbolic
system the distribution of these trapping times is exponential,
and can be characterized by a mean trapping time $\gamma^{-1}$ or its
inverse, the escape rate $\gamma$. One method to calculate $\gamma$ makes use
of trapped periodic orbits. Roughly speaking, if a trajectory makes $n$
bounces, then it will be close to some periodic orbit of that period. The
orbit is exponentially unstable and the probability of staying near it is
given by the inverse of the expanding eigenvalue. The average over all
periodic points, including multiple traversals, then is
\beq
\sum_{p} \sum_{r=1}^\infty
 T_p  |\Lambda_p|^{-r} e^{- T_p s r } \, ,
\eeq
which is exactly of the form of the semiclassical expression above, with
the exception that now a full eigenvalue appears rather than its square
root. Therefore, by manipuilations similar to the ones in eq (\ref{z_deriv}),
one can write this a the logarithmic derivative of the zeta function
\beq
1/\zeta = \prod_{p}
 \left( { 1 - |\Lambda_p|^{-1} e^{-T_p s} } \right) \, .
\label{z_es}
\eeq
Using cycle expansions up to $n=1$ and $n=2$, respectively, one finds
the curves shown in Fig.~\ref{3-d_esc}.
The agreement between the two curves is an indication of the
rapid convergence of the cycle expansion. Also shown are the
results of a classical Monte Carlo simulation of Gaspard\cite{gaspard1}.

There is a direct connection between the escape rate\cite{TK} and the
classical sum rule proposed by Hannay and Ozorio de Almeida\cite{OdAH}.
According to one formulation (see in particular section 17.7 of
Gutzwiller's book\cite{Gutz90}), the sum
\beq
\sum_{T_p<T} {T_p \over \det(1-M_p)} \approx 1 \, ,
\eeq
so that the set of periodic orbits approximates the phase space measure
in the sense that every average over phase space can be expressed as
an average over periodic orbits. Rather than defining a set of orbits by their
periods, one can also take all periodic orbits of a fixed symbol length. Then
the equivalent statement is that
\beq
\sum_{(n)} {T_p \over \det(1-M_p)} \approx 1 \, .
\eeq
This may be expressed as a logarithmic derivative of a zeta function
by multiplication with $z^n$ and summation over $n$. That the sum approaches a
constant then means that the zeta function
\beq
1/\zeta = \prod_p (1-|\Lambda_p|^{-1} z^{n_p})
\label{sum_rule}
\eeq
has a zero for $z=1$. But this is just the escape rate zeta function
eq~(\ref{z_es}) for $s=0$, so that a vanishing escape rate indicates
that the periodic orbits represent the invariant phase space
density (in hyperbolic systems).

\section{Matrix elements}
In this final section, I will discuss the derivation of a zeta function
for the diagonal matrix elements and indicate the extension to
off-diagonal matrix elements\cite{EFMW91}.

\subsection{Diagonal matrix elements}

As discussed in section 2.1, the final expression for $\tr GA$ with some
sufficiently smooth observable $A$ becomes:
\bea
\rho_A(E) &=& \int { d{\bf q} d{\bf p} \over h^N} A({\bf q}, {\bf p})
           \delta(E-H({\bf q}, {\bf p})) \,                             \cr
 &\  & + {\rm Im\,} \> {i \over \pi \hbar}
\sum_{p} \sum_{r=1}^\infty { A_p  \over | \det (M_p^r-1) |^{1/2}  }
e^{(i S_p(E)/\hbar - i \pi \mu /2)r} \, ,
\eea
where
\beq
A_p = \int_0^{T_p} dt\, A({\bf q}_p(t), {\bf p}_p(t)) \, .
\eeq

The derivation of a zeta function for matrix elements is similar
to the calculation in section 2.2, except for one step.
One first arrives at
\beq
R_{A,osc} = {1 \over \pi \hbar}
\sum_p \sum_{j=0}^\infty { A_p \tilde{t}_p^{(j)} \over
1- \tilde{t}_p^{(j)} } \, ,
\eeq
where  $\tilde{t}_p^{(j)} =
e^{i S_p/\hbar - i\mu_p\pi/2} |\Lambda_p|^{-1/2} \Lambda_p^{-j}$.
In the case of the density of states, the observable is $A=1$ and the
prefactor $A_p=T_p = \partial S / \partial E$ so that one can
immediately write the quotient in~(\ref{log_der}) as a logarithmic
derivative and thus arrive at the zeta functions. Here, one may proceed
similarly after introducing an auxiliary variable $\gamma$ and an
extended $ t_p^{(j)} = e^{-\gamma A_p} \tilde{t}_p^{(j)}$.
Since $\tilde{t}_p^{(j)} = t_p^j(\gamma = 0)$ and
$A_p \tilde{t}_p^{(j)} = \partial t_p^{(j)} / \partial \gamma |_{\gamma = 0}$,
one can write
\beq
R_{A,osc} = {i \over \pi \hbar}
\sum_p \sum_{j=0}^\infty {\partial \over \partial \gamma}
\log \left. (1-t_p^{(j)})\right|_{\gamma=0}
\eeq
and thus finally
\beq
R_{A,osc} = {i \over \pi \hbar}
\left. {\partial \over \partial \gamma} {\rm ln} Z(\gamma,E)
\right|_{\gamma=0}
\eeq
with the extended Selberg type product
\beq
Z(\gamma, E) = \prod_p \prod_{j=0}^\infty (1-e^{-\gamma A_p}
 e^{i S_p /\hbar- i \mu_p \pi / 2} |\Lambda_p|^{-1/2} \Lambda_p^{-j} ).
\label{Z_gam}
\eeq
For $\gamma=0$ this dynamical zeta function coincides with the Selberg Zeta
function $Z_s(E)$ for the spectral density $\rho(E)$,
\beq
Z_s(E) = Z(\gamma=0, E)\, .
\eeq

Eq~(\ref{Z_gam}) may again be evaluated using the cycle
expansion. Consider the case of a hyperbolic system with
complete binary symbolic dynamics. Labelling
the contributions from the periodic orbits by their symbolic code, one
has to consider products of the form
\beq
Z(\gamma, E) = (1-t_0) (1-t_1) (1-t_{01})(1-t_{001}) (1-t_{011}) \cdots
\, ,
\eeq
which expand into
\beq
Z(\gamma, E) = 1- t_0 - t_1 - (t_{01} - t_0 t_1)
- (t_{001} - t_0 t_{01}) - (t_{011} - t_1 t_{01}) \cdots
\eeq
Upon taking the derivative, one has
\bea
{\partial Z(\gamma, E) \over \partial \gamma}
& = &A_0 t_0 + A_1 t_1 + \lbrack A_{01} t_{01} - (A_0+A_1) t_0 t_1\rbrack
 \cr
& & + \lbrack A_{001} t_{001} - (A_0 + A_{01}) t_0 t_{01}\rbrack  \cr
& & + \lbrack A_{011} t_{011} - (A_1 + A_{01}) t_1 t_{01}\rbrack \cdots
\eea
The suggestive argument for convergence (section 5.4)
is that a long orbit $t_{ab}$
can be decomposed into shorter ones (labelled $a$ and $b$),
which shadow it; therefore, the term $t_{ab} - t_a t_b$ is small.
However, the same can be expected of the observable: if the short
orbits $a$ and $b$ are close to $ab$, then $A_{ab}$ should also be
close to $A_a + A_b$ and the convergence properties of the above
cycle expansion can be expected to be the same as the ones for the
density of states.

\subsection{Correlation functions}

A similar result may be obtained for correlation functions.
All one has to note is that in the semiclassical limit the
time evolution of the operator $B$ is given by the classical
observable $B$ at the time evolved positions. Therefore,
one obtains correlation functions if the observable $A({\bf p}, {\bf q})$ is
replaced by $B({\bf p}(t), {\bf q}(t)) C({\bf p}, {\bf q})$, where
$({\bf p}(t), {\bf q}(t))$ denotes the point reached by a trajectory starting
at time $t=0$ from $({\bf p}, {\bf q})$.
The final result is exactly as above, except that $A_p$ is replaced by
\beq
A'_p = \int_0^{T_p} dt' B({\bf q}(t'+t), {\bf p}(t'+t))
C({\bf q}(t'), {\bf p}(t')) \, ,
\label{BCf}
\eeq
the correlation function of $B$ and $C$ along the periodic orbit.
The first term, the contribution of paths of zero length,
becomes the classical correlation function.

Quantum mechanically, $\tr\, G\, B_t\,C$ evaluates to
\beq
\tr\, G\, B_t\,C = \sum_n {\langle n | B_t C | n \rangle \over E - E_n}
\, ,
\eeq
where
\beq
\langle n | B_t C | n \rangle = \sum_m
\langle n | B | m \rangle\,  \langle m | C | n \rangle
e^{i (E_n-E_m)t /\hbar};
\eeq
one therefore expects the semiclassical expression to have poles
at the eigenenergies, with residues the matrix elements
$\langle n | B_t C | n \rangle$.

\subsection{Periodic orbit spectroscopy}

The developments in the previous section pave the way for what one
may call `periodic orbit spectroscopy'\cite{Comm}. A Fourier
transform of spectra in terms of a suitable scaling variable very often
reveals sharply defined peaks, which according to semiclassics
can be assigned to periodic orbits. The most prominent example
of this is hydrogen in a magnetic field, where the comparison
between classical and quantum mechanics has been pushed very
far\cite{Wel,Wi87,DuDel}.

The formula for autocorrelation functions should be useful
in describing Rydberg-Rydberg transitions. Calculations by Zeller\cite{Zel}
for hydrogen in a magnetic field have in fact revealed periodic orbit
structures for transitions between highly excited states. A quantitative
comparison  would be highly desirable.

As an application outside atomic physics,  Wilkinson\cite{Wil2}
has suggested to apply the above formula to conductance problems,
since the conductance is related to the Fourier transform of a correlation
function by the Kubo Greenwood formula\cite{Kubo,Gre}.
This relation may be useful for the AC behaviour, but must be
expected to fail in the DC limit since it does not describe
correctly the infinite time limit of the correlation functions\cite{EFMW91}.

Periodic orbits also figure prominently in recent interpretations of the
spectra of small molecules\cite{GP}.  Their presence becomes plausible via
the classical periodic orbit formula. It would be interesting to
identify semiclassical corrections to this.

\section{Final remarks}

Evidently, periodic orbit expansions provide a powerful tool for all kinds
of calculations in classical and semiclassical mechanics. Almost all
quantities of interest can be expressed in terms of periodic orbits or,
more precisely, traces of transfer matrices. The cycle expansion with its
arrangement of periodic orbits into compensating groups provides a
numerically efficient and convergent method
to evaluate periodic orbit expressions.

The tests presented here and others have aimed at verifying the
semiclassical approximation level by level. This is mainly of theoretical
interest, to strengthen confidence and explore the range of validity
and reliability of the Gutzwiller trace formula, which after all is derived
using many approximations. The indications are that it is a bona fide
semiclassical theory, which shows large deviations from exact quantum
calculation in the quantum domain of small wavenumbers, but better
agreement at higher energies in the truly semiclassical regime.
There still is the question of whether the
present form of semiclassics can actually give the positions of resonances
accurately to better than a mean spacing, considering that the
propagator includes terms of order $\hbar$ only, whereas the mean density
of states is $O(\hbar^2)$. Indications are that the error will not be
noticable until one gets to very high energies. However, it should be
more important as the number of degrees of freedom increases.

On the practical side, this appears to be just a minor limitation. The
strength of the semiclassical expressions is that they can explain long range
correlations in the spectra, somewhat independent of the
fince structure of the spectrum. The aim will be to calculate
semiclassically from just a few periodic orbits the spectra of small molecules
and to use this as a tool in analyzing spectra. The most prominent example of
this class is hydrogen in a magnetic field, but currently some molecules are
under investigation where such quantitative predictions may be possible
as well.

\section*{Acknowledgements}

It is a pleasure to thank E.B. Bogomolny,
P. Cvitanovi\'c, S. Fishman,
K. M\"uller, G. Russberg, P. Scherer and
D. Wintgen for stimulating discussions and fruitful collaboration.
Critical comments on the manuscript by G. Russberg and
D. Wintgen are greatfully acknowledged.

\newpage

\newcommand{\AP}[1]{{\em Ann.\ Phys. (NY)}\/ {\bf #1}}
\newcommand{\CMP}[1]{{\em Commun.\ Math.\ Phys.}\/ {\bf #1}}
\newcommand{\JCP}[1]{{\em J.\ Chem.\ Phys.}\/ {\bf #1}}
\newcommand{\JETP}[1]{{\em Sov.\ Phys.\ JETP}\/ {\bf #1}}
\newcommand{\JETPL}[1]{{\em JETP Lett.\ }\/ {\bf #1}}
\newcommand{\JMP}[1]{{\em J.\ Math.\ Phys.}\/ {\bf #1}}
\newcommand{\JMPA}[1]{{\em J.\ Math.\ Pure Appl.}\/ {\bf #1}}
\newcommand{\JPA}[1]{{\em J.\ Phys. A: Math. Gen. }\/ {\bf #1}}
\newcommand{\JPB}[1]{{\em J.\ Phys. B: At. Mol. Opt. }\/ {\bf  #1}}
\newcommand{\JPC}[1]{{\em J.\ Phys.\ Chem.}\/ {\bf #1}}
\newcommand{\PLA}[1]{{\em Phys.\ Lett.}\/ {\bf A #1}}
\newcommand{\PRA}[1]{{\em Phys.\ Rev.}\/ {\bf A #1}}
\newcommand{\PRL}[1]{{\em Phys.\ Rev.\ Lett.}\/ {\bf #1}}
\newcommand{\PST}[1]{{\em Phys.\ Scripta }\/ {\bf T #1}}
\newcommand{\RMS}[1]{{\em Russ.\ Math.\ Surv.}\/ {\bf #1}}
\newcommand{\USSR}[1]{{\em Math.\ USSR.\ Sb.}\/ {\bf #1}}
\renewcommand{\baselinestretch} {1}

\newpage

\small{   %begin small

\begin{tabular}{|r|c|rrr|rr|}
\hline
n  & $M_n(N)$& $M_n(2)$& $M_n(3)$& $M_n(4)$
                            %   &   $M_n^{3-disk}$& $M_n^{4-disk}$
\cr \hline
    1  &   N                     &  2 &    3 &      4 %& 0 & 0
\cr 2  &$N (N-1)/2$              &  1 &    3 &      6 %& 3 & 6
\cr 3  &$N (N^2-1)/3$            &  2 &    8 &     20 %& 2 & 8
\cr 4  &$N^2(N^2-1)/4$           &  3 &   18 &     60 %& 3 & 18
\cr 5  &$(N^5-N)/5$              &  6 &   48 &    204 %& 6 & 48
\cr 6  &$(N^6-N^3-N^2+N)/ 6$     &  9 &  116 &    670 %& 9 & 116
\cr 7  &$(N^7-N)/ 7$             & 18 &  312 &   2340 %& 18 & 312
\cr 8  &$N^4(N^4-1)/ 8$          & 30 &  810 &   8160 %& 30 & 810
\cr 9  &$N^3(N^6-1)/ 9$          & 56 & 2184 &  29120 %& 56 & 2184
\cr 10 &$(N^{10}-N^5-N^2+ N)/ 10$& 99 & 5880 & 104754 %& 99 & 5880
\cr \hline
\end{tabular}
\vskip .3cm

Table 1. Number of prime cycles for various alphabets and grammars up to
length 10. The first column gives the cycle length, the second the formula
(\ref{nppo}) for the number of prime cycles for complete $N$-symbol
dynamics, columns three through five give the numbers for $N=2, 3$ and $4$.

\bigskip

\begin{tabular}{|l|l|l|}
\hline
$\tilde{p}$ & ${p}$  & ${\bf g}_{\tilde{p}}$ \\
\hline
0  &  1\,2  &  ${\sigma}_{12}$ \\
1  &  1\,2\,3  &  $C_3$ \\
\hline
01  &  12\,13  &  ${\sigma}_{23}$ \\
\hline
001  &  121\,232\,313  &  $C_3$ \\
011  &  121\,323  &  ${\sigma}_{13}$ \\
\hline
0001  &  1212\,1313  &  ${\sigma}_{23}$ \\
0011  &  1212\,3131\,2323  &  $C_3^2$ \\
0111  &  1213\,2123  &  ${\sigma}_{12}$ \\
\hline
00001  &  12121\,23232\,31313  &  $C_3$ \\
00011  &  12121\,32323  &  ${\sigma}_{13}$ \\
00101  &  12123\,21213  &  ${\sigma}_{12}$ \\
00111  &  12123  &  $e$ \\
01011  &  12131\,23212\,31323  &  $C_3$ \\
01111  &  12132\,13123  &  ${\sigma}_{23}$ \\
\hline
000001  &  121212\,131313  &  ${\sigma}_{23}$ \\
000011  &  121212\,313131\,232323  &  $C_3^2$ \\
000101  &  121213  &  $e$ \\
000111  &  121213\,212123  &  ${\sigma}_{12}$ \\
001011  &  121232\,131323  &  ${\sigma}_{23}$ \\
001101  &  121231\,323213  &  ${\sigma}_{13}$ \\
001111  &  121231\,232312\,313123  &  $C_3$ \\
010111  &  121312\,313231\,232123  &  $C_3^2$ \\
011111  &  121321\,323123  &  ${\sigma}_{13}$ \\
\hline
\end{tabular}

\vskip .5cm

Table 2.
$C_{3v}$ correspondence between the binary labelled fundamental domain prime
cycles $\tilde{p}$ and the full 3-disk ternary \{1,2,3\} cycles ${p}$,
together with the $C_{3v}$ transformation that maps the end point of the
$\tilde{p}$ cycle into the irreducible segment of the $p$ cycle. The degeneracy
of $p$ cycle is $m_p=6 n_{\tilde{p}}/n_p$. The shortest
pair of the fundamental domain prime cycles related by time symmetry are the
6-cycles $\overline{001011}$ and $\overline{001101}$.
\vfill\eject

} % end small

{\bf {Figure captions}}
\vskip 20pt

\begin{enumerate}

\fg{traj}
The three disk billiard. The trajectory can be labelled
by its disk visitation sequence, viz. $1231312$.

\fg{s_int}
Construction of impact parameters for long collision sequences.
The dashed lines indicate the interval in impact parameter leading
to collisions with disk 1. Initial conditions in the two shaded intervals
lead to collisions with 1 and 2 or 1 and 3, respectively.

\fg{orbits}
Two orbits for hydrogen in a magnetic field in semiparabolic
coordinates. The energy is $+0.2\,Ry$. The self conjugate points
are indicated. The four disk code of the two trajectories
is $142124$ and $14{\bf 1}212{\bf 1}4$, respectively. The
collisions indicated in bold face are not immediately obvious in
configuration space but clearly identified by the presence of two additional
self conjugate points.

\fg{3d_symm}
Symmetries of the three disk scattering system. Indicated are the
three reflections $ \sigma_i$ across symmetry lines,  the two
rotations $C_3$ and $C_3^2$ around the center by $\pm 2\pi/3$
and the fundamental domain (shaded).

\fg{3-d_conv}
The abscissa of absolute convergence for the system of three disks
as a function of the distance/radius ratio $d/R$.

\fg{3-d-res}
Some resonances of the $S$-matrix in the $A_1$ subspace for $d/R=6$.
The crosses are the exact quantum results, the open circles the result
from a semiclassical calculation involving all periodic orbits
up to symbol length 5. Notice that all semiclassical
resonances lie below the abscissa of absolute convergence,
${\rm Im\,} k_c = -0.123...$, but that the first two quantum
resonances lie above.

\fg{cycle_conv}
Convergence of the cycle expanded zeta function for the three disk
system. The terms $c_n$ contain all contributions from orbits and
pseudo-orbits of length $n$ (see eq~(\ref{curv_n})) at $k=0$.

\fg{DE}
The functional determinant (\ref{f_det}) for the closed
three disk system including orbits of symbol length $n=2$ and
$n=3$. The dotter vertical lines
indicate the positions of the exact quantum eigenvalues.

\fg{3-d_esc}
Classical escape rate $\gamma$ as a function of the distance between the
disks. The crosses are the results of a Monte Carlo simulation
by Gaspard\cite{gaspard1}.
%
%\fg{pr}
%The classical sum rule (\ref{sum_rule}). Shown
%are the leading zeros $z_0$ of the cycle expanded zeta function
%(\ref{sum_rule}) for increasing cycle length $n$.

\end{enumerate}

\end{document}